\documentclass[twocolumn,showpacs,preprintnumbers,amsmath,amssymb]{revtex4}
\usepackage{tabularx,graphicx}\begin{document}
\newcommand{\beq}{\begin{equation}}
\newcommand{\eeq}{\end{equation}}
\newcommand{\beqn}{\begin{eqnarray}}
\newcommand{\eeqn}{\end{eqnarray}}
\newcommand{\bmath}{\begin{subequations}}
\newcommand{\emath}{\end{subequations}}

\title{ Electron-hole asymmetry is the key to superconductivity}
\author{J.E. Hirsch}
\address{Department of Physics, University of California, San Diego\\
La Jolla, CA 92093-0319}

\date{January 31, 2003} 

\begin{abstract}
In a solid, transport of electricity can occur via negative electrons or via positive holes. In the normal state of superconducting materials
experiments show that transport is  usually dominated by $dressed$ $positive$ $hole$ $carriers$. Instead, in the superconducting state experiments show that the supercurrent
is always carried by $undressed$ $negative$ $electron$ $carriers$. These experimental facts indicate that  electron-hole
asymmetry plays a fundamental role in superconductivity, as proposed by the theory of hole superconductivity.
\end{abstract}
\maketitle

\section{Holes are not like electrons}	

To understand superconductivity in solids it is essential to understand the fundamental electron-hole asymmetry of
condensed matter. This asymmetry originates in the fact that the positive proton is 2000 times heavier than the negative
electron.

When an electronic energy band is almost empty the carriers are electrons, when the band is almost full the carriers
 are holes. Ever since Heisenberg\cite{heisenberg} introduced the concept of 'hole' in 1931 for the description of atomic spectra, electrons and holes
have been regarded as equivalent quasiparticles in solids, as described for example by Peierls\cite{peierls}:
``Ein Band, in dem sich nur wenige Elektronen befinden, verhalt sich in jeder Beziehung genau so, wie ein Band in dem
nur fur wenige Elektronen noch Platz ist''. We assert that 'in jeder Beziehung' is certainly wrong, and that in fact
superconductivity originates in the fundamental asymmetry between electrons and holes in condensed matter.

\begin{figure}
\resizebox{8cm}{!}{\includegraphics[width=9cm]{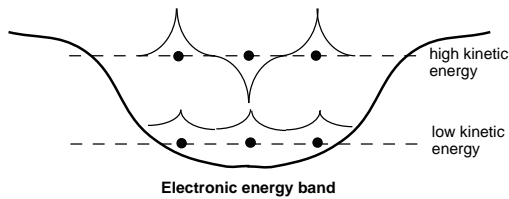}}
\caption{Electronic energy states in a band (schematic). The states near the bottom (top)  of the band are bonding (antibonding).}
\label{fig1}
\end{figure}

How are electrons and holes different? Figure 1 depicts schematically an electronic energy band and the 
wavefunction of the carriers at the Fermi level when the Fermi level is near the bottom and near the top of the
band. The electrons near the bottom of the band have a $bonding$ wave function, with large amplitude in the
region between the ions. The electrons near the top of the band have an $antibonding$ wave function, with the
wave function having a node in the region between the ions. This difference is independent of the nature of the
orbitals giving rise to the bands: it originates in the fact that the states at the bottom of the band choose their
wavefunction so as to minimize their kinetic plus electron-ion energy, and that the states at the top of the band are
forced by the Pauli principle to have their wavefunction orthogonal to all the lower states in the band.

The antibonding electrons near the top of a band are $not$ a happy bunch. Because of the large oscillations in their
wave function forced by having to be orthogonal to the bonding electrons, they have a high kinetic energy.
Also,  their wavefunction is such that it exerts 'negative pressure' that tends to 
 break the solid apart (hence their name, antibonding), unlike the bonding electrons near the bottom of the band
that tend to bind the solid. Furthermore,  when an external electric
field is applied they move in direction $opposite$ to the applied force, countering rather than helping the
transport of electricity done by the bonding electrons. Moreover, they are seldom talked about:  when the Fermi level is 
near the top of the band,
it is conventional to describe transport by the time evolution of the few empty states in the band (holes) rather
than to describe the opposite contributions of bonding and antibonding electrons; in such case it is said
that transport is done by holes rather than electrons. 

There are other fundamental differences between the carriers at the Fermi energy when the Fermi level is near the
bottom and near the top of a band. In a given band, electron carriers are always less dressed than hole carriers\cite{holeelec,dynhelec}.
Because the electron-electron repulsion is always bigger than the separation between electronic energy
levels, when a second electron occupies a Wannier orbital already occupied by an electron the state of the
first electron is modified, causing a smaller quasiparticle weight and a larger effective
mass for hole carriers compared to electron carriers\cite{dynh}.

\begin{table}[h]
\caption{Electrons versus holes at the Fermi energy}
{\begin{tabular}{@{}ll@{}} \toprule
{\bf Electrons} & {\bf Holes}  \\
 \colrule
Bonding states & Antibonding states\\
Low kinetic energy&High kinetic energy\\
Small effective mass&Large effective mass\\
Negatively charged carriers&Positively charged carriers\\
Large quasiparticle weight& Small quasiparticle weight \\
Coherent conduction&Incoherent conduction \\
Large Drude weight&Small Drude weight \\
Good metals& Bad metals \\ 
Stable lattices& Unstable lattices \\ 
Ions attract each other& Ions repel each other \\ 
Carriers repel each other& Carriers attract each other \\ 
Normal metals& Superconductors  \\ 
\botrule
\end{tabular}}
\end{table}

Table 1 summarizes the differences between carriers at the Fermi energy when the Fermi enery is near the
bottom and near the top of the band. It is clear that holes are not like electrons.
It is natural to expect that these differences can have fundamental consequences for superconductivity. The single
band Hubbard model, widely used to describe interacting electrons in energy bands, is electron-hole symmetric and hence
it is inadequate to describe the physics of real electrons in energy bands in solids. 'Dynamic Hubbard models' are
required instead\ \cite{holeelec,dynhelec,dynh}.

\section{Electron-hole asymmetry and superconductivity}

Having established in the previous section that holes are different from electrons, we
assert that superconductivity originates in the fundamental asymmetry beween electrons and holes, as follows:

(1) Superconductivity can only occur when carriers in the normal state are holes.

(2) Holes undress and become like electrons when they pair.

(3) The superfluid carriers are electrons.

In the following we discuss these points in more detail.

\subsection{The carriers of electricity in the normal state  are dressed holes}
Since the early days of superconductivity it was noted that if a metal does not have hole carriers it does not
become superconducting, hence that the presence of hole carriers is a $necessary$ condition for superconductivity
\cite{kikoin,born,chapnik}.
Furthermore non-superconductors are usually better conductors of electricity than superconductors in the
normal state. We attribute the poor normal state  conductivity of superconductors to the fact that hole carriers are
'dressed' due to interactions and hence their effective mass is higher and correspondingly they give a 
smaller contribution to the low frequency conductivity. Optical experiments in high $T_c$ cuprates clearly
show that the intraband Drude weight is small and there is large optical absorption in the mid-intrared and
higher frequency range\cite{uchida}. Photoemission experiments in the cuprates in the 
normal state do not show clear evidence
for quasiparticles except in the overdoped regime, indicating that carriers are heavily dressed
\cite{photonorm}.

\subsection{Superconductivity is driven by hole undressing}\
\begin{figure}
\resizebox{8cm}{!}{\includegraphics[width=9cm]{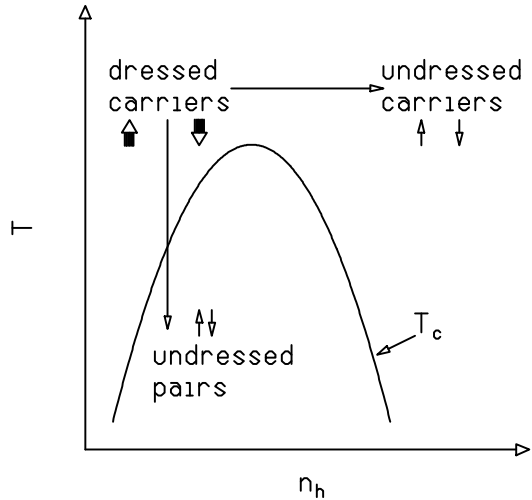}}
\caption{Phenomenology of hole superconductivity. When the local hole concentration increases
either through doping or through pairing, holes undress. From its maximum, $T_c$ decreases
to the left as the number of holes goes to zero and to the right as the holes undress in the
normal state and no longer need to pair. }
\label{fig2}
\end{figure}

In the theory of hole superconductivity, dressed hole carriers undress when the 'local' hole concentration around a given carrier increases\cite{undressing}. This will happen both when the hole concentration increases in the normal state by doping, and
when hole carriers pair as the temperature is lowered. Undressing leads to larger quasiparticle weight and smaller 
effective mass of the carriers, and the resulting lowering of kinetic energy provides the condensation energy
for the superconducting state\cite{kine}. The phenomenology is shown in Figures 2 and 3 and was predicted\cite{color}
many years before it became clear from recent optical\cite{optical}, photoemission\cite{photo}
 and transport\cite{transport} experiments.
\begin{figure}
\resizebox{8cm}{!}{\includegraphics[width=9cm]{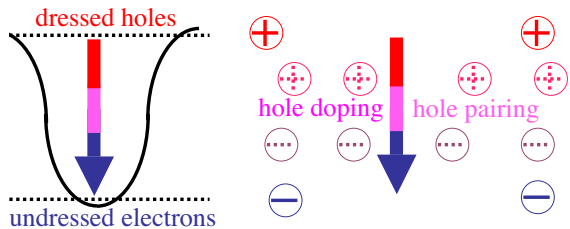}}
\caption{Phenomenology of hole superconductivity. As the Fermi level goes up
in the band, bare electrons become dressed holes. When holes pair it is as if locally 
the band becomes less full, hence holes undress and turn into electrons.}
\label{fig3}
\end{figure}

\subsection{The superfluid carriers in the superconducting state are undressed electrons}
\begin{figure}
\resizebox{9.5cm}{!}{\includegraphics[width=7cm]{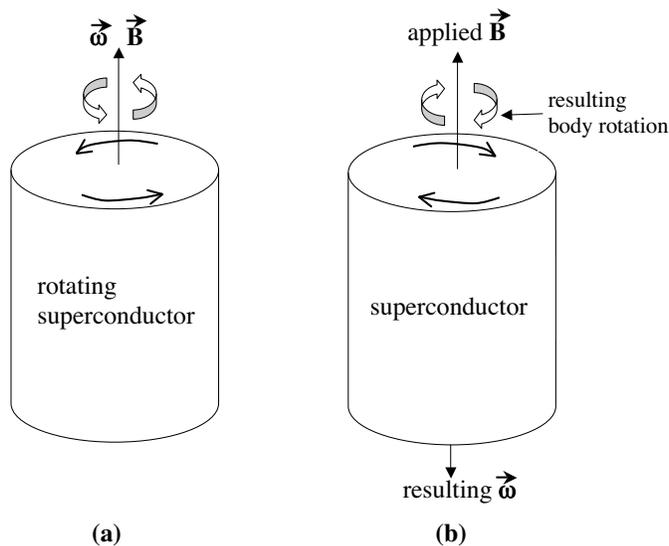}}
\caption{(a) A superconductor rotating counterclockwise as seen from the top gives rise to a magnetic field pointing up,
parallel to $\vec{\omega}$, if the mobile carriers in the superconductor have $negative$ charge.
(b) When a magnetic field is suddenly turned on pointing $up$ the superconductor will start to rotate $clockwise$ as seen
from the top, i.e. with angular velocity $antiparallel$ to the applied field, if the mobile carriers have $negative$ charge. }
\label{fig4}
\end{figure}

The nature of superfluid carriers in the superconducting state is revealed by experiments where a 
superconductor is put into rotation. The London equation implies that a magnetic field is generated (London field), given by\cite{london}
\beq
\vec{B}=-\frac{2mc}{q}\vec{\omega}
\eeq
where $q$ and $m$ are the charge and mass of the superfluid carriers , $c$ is the speed of light 
and $\vec{\omega}$ the angular velocity. The field is generated because the superfluid 'lags behind' the 
solid rotation in a surface layer of thickness given by the 
penetration depth. Experimentally it is determined\cite{londonexp} that the charge in Eq. (1) is the electronic charge $with$ $its$ $sign$,
i.e. that the generated magnetic field is parallel to the angular velocity, and that $m$ is the $bare$ electron mass. These
experiments, as well as related experiments where a superconductor is put into rotation by application of
a magnetic field\cite{gyro}, reveal that the superfluid carriers are bare electrons with negative charge. In other words,
the dressed positive hole carriers in the normal state turn into undressed negative electron carriers in 
the superconducting state.

\section{Discussion}
It is impossible to give a comprehensive description of the theory of hole superconductivity
in this short paper.  A recent more extended review is given in ref. 20.
As experiments improve, more of the predictions of the theory made early on become
experimental facts. For example, the tunneling asymmetry predicted back in 
1989\cite{tunneling}
has become an experimental fact many years later; the strong to weak coupling crossover
with increasing hole concentration predicted in 1989\cite{strongweak} has only
very recently been confirmed\cite{hc2}. The optical sum rule violation and color change predicted
back in 1992\cite{color} has been verified experimentally recently\cite{optical}. The prediction that high
$T_c$ will result whenever holes in an almost filled band conduct through
negatively charged ions\cite{ions} is confirmed by the discovery
of 40K superconductivity in $MgB_2$\cite{mgb2}. The prediction\cite{twoband} that when holes and electrons are present at the  Fermi level holes develop a large gap and electrons a small gap is confirmed by recent
tunneling experiments in $MgB_2$\cite{twogaps}. More examples will follow.

\end{document}